\documentclass[twoside]{JHEP3mod}

\usepackage{graphicx}
\usepackage{amsmath}

\def\dsl{\raise.15ex\hbox{/} \kern-.57em\partial}

\title{Finite Density QCD: a New Approach}

\author{Vicente Azcoiti \\ 
Departamento de F\'{\i}sica Te\'orica, Universidad de
Zaragoza, Cl. Pedro Cerbuna 12, E-50009 Zaragoza (Spain) \\
E-mail: \email{azcoiti@azcoiti.unizar.es}} 
\author{Giuseppe Di~Carlo \\ 
INFN, Laboratori Nazionali del Gran Sasso, 
67010 Assergi,(L'Aquila) (Italy) \\
E-mail: \email{gdicarlo@lngs.infn.it}} 
\author{Angelo Galante \\
INFN, Laboratori Nazionali del Gran Sasso, 
67010 Assergi,(L'Aquila) (Italy) and
Dipartimento di Fisica dell'Universit\`a di L'Aquila,
67100 L'Aquila (Italy) \\
E-mail: \email{galante@lngs.infn.it}}
\author{V\'{\i}ctor Laliena \\
Departamento de F\'{\i}sica Te\'orica, Universidad de
Zaragoza, Cl. Pedro Cerbuna 12, E-50009 Zaragoza (Spain) \\
E-mail: \email{laliena@unizar.es}}

\abstract{
We introduce a new approach to analyze the phase diagram of QCD at finite 
chemical potential and temperature, test it in the Gross-Neveu model at 
finite baryon density, and apply it to the study of the chemical 
potential-temperature phase diagram of QCD with four degenerate flavors of 
Kogut-Susskind type.}


\begin{document}

\section{Introduction}

Hadronic matter at high density and finite temperature is expected to 
undergo one or more phase transitions from a confined phase to a quark-gluon 
plasma phase, and eventually a color superconducting phase 
due to a (weak) attractive  
channel between quarks of different colors, that could lead to the
formation of quark pairs, analogous of Cooper pairing in solid state 
systems at low temperature. 
The behavior of hadronic matter at high density is also relevant 
in astrophysics, for the study of neutron stars
and extremely compact objects, and in cosmology, for the understanding of
phase transitions in the early Universe. Furthermore 
the new states of matter should be partially accessible to direct
experimental verification in the heavy ion collision experiments.

These features of finite density QCD can be studied only
nonperturbatively. Unfortunately the lattice approach, 
the most powerful tool to perform
first principles, nonperturbative studies, 
is affected in the case of finite density QCD by the well known 
sign problem, that has prevented for a long time any step toward its
understanding. Indeed the moderate optimism generated by the finding of a 
first order phase transition at high density in the strong-coupling limit 
\cite{barbour} using the Glasgow reweighting method, was actually ill-founded 
\cite{noi}.

From the accumulated experience it was clear that the only way to minimize the 
effect of the phase of the fermion determinant in the integration measure 
of finite density QCD would be to work in the limit of large fermion masses. 
In fact taking the limit of fat QCD and using the MFA approach \cite{mfa} in 
the numerical simulations, we were able to predict a tentative phase diagram 
for fat QCD \cite{noi3} which qualitatively confirmed the theoretical 
expectations.

This was the rather unpleasant situation of three-color QCD at finite density 
until two years ago, when Philippe de 
Forcrand and Owe Philipsen \cite{philippe}, and later on D'Elia and 
Lombardo \cite{mp}, reported some reliable results for light quarks, 
obtained from simulations at imaginary chemical potential 
where the sign problem, as well known, is absent. Fodor and Katz have also 
reported interesting results \cite{fodor} 
obtained within a modified version in two-parameter space of the Glasgow 
re-weighting technique \cite{barbour} which, due to the 
re-weighting procedure, can be applied only to relatively small lattice sizes.
A modified version of the last approach, which makes use of a Taylor expansion 
in the chemical potential, has produced also interesting results for larger 
lattice sizes in the small $\mu/T$ region \cite{karsch}.

This paper is devoted to the presentation of a new approach to simulate QCD 
at finite temperature and baryon density, and to the application of this 
approach to the four-fermion model in three dimensions and to QCD with four 
degenerate quarks. Even if our theory scheme can resemble in some aspects 
the imaginary chemical potential approach, its range of applicability is 
at first sight
much wider, as will become apparent after reading section 2, which we devote 
to a detailed exposition of the approach. Section 3 contains the results of a 
test worked on the three-dimensional Gross-Neveu model at finite baryon 
density and in the large number of flavors limit. This model shows a 
non trivial phase diagram in the $T, \mu$ plane \cite{sacha} and in 
addition, in the large N limit, can be analytically solved, thus allowing 
to confront results obtained within our approach with exact ones. In section 
4 we show our results for the phase diagram of QCD at finite chemical 
potential with four degenerate flavors of Kogut-Susskind type. The data 
reported in this section were obtained from simulations at values 
of the gauge coupling and fermion masses in the scaling region. Last section 
is devoted to collect our main conclusions and to discuss possible issues 
for near future.

\section{Theory scheme}

The lattice action for QCD at finite chemical potential $\mu$ and staggered 
fermions is 

\begin{eqnarray}
S &=& S_{\mathrm PG} + \frac{1}{2}\sum_n\sum^3_{i=1}
\bar\psi_n \eta_i (n) \left( U_{n,i}
\psi_{n+i} - U^\dagger_{n-i,i}\psi_{n-i}\right) \nonumber \\
&+& \frac{1}{2}\sum_{n} \bar\psi_n \eta_0 (n) \left( e^\mu  U_{n,0}\psi_{n+0} 
- e^{-\mu} U^\dagger_{n-0,0}\psi_{n-0}\right) + m \sum_n \bar\psi_n\psi_n
\label{action}
\end{eqnarray}

\noindent
where $S_{\mathrm PG}$ is the standard Wilson action for the gluonic fields 
and $\eta_\mu (n)$ the Kogut-Susskind phases. 
The temporal part of the fermionic 
action can also be written as

\begin{eqnarray}
S_{\tau} &=& \cosh(\mu) \frac{1}{2} 
\sum_{n}\bar\psi_n \eta_0 (n) \left( U_{n,0}
\psi_{n+0} - U^\dagger_{n-0,0}\psi_{n-0}\right) \nonumber \\
&+& \sinh(\mu) \frac{1}{2}  \sum_{n} \bar\psi_n \eta_0 (n)\left( U_{n,0}
\psi_{n+0} + U^\dagger_{n-0,0}\psi_{n-0}\right)
\label{actiont}
\end{eqnarray}

The contribution proportional to $\cosh(\mu)$ in $S_\tau$ is a fermionic 
bilinear
defined through a matrix $i\Lambda_\tau$ with $\Lambda_\tau$ 
a hermitian matrix which anticommutes with $\gamma_5$. 
The contribution proportional to $\sinh(\mu)$ is also a fermionic bilinear 
defined 
through the hermitian matrix $\bar\Lambda_\tau$. Since the space-like part 
of the fermionic action is also a bilinear defined by the matrix 
$m I + i \Lambda_s$, with  $\Lambda_s$ hermitian, it is evident that the 
introduction of a chemical potential $\mu$ drive us to a complex fermion 
effective action. At the same time it is also obvious that an imaginary 
chemical potential solves the sign problem since $\cosh(\mu)$ becomes the 
trigonometric cosine and $\sinh(\mu)$ the trigonometric sine, but the last 
picking up a factor of i.

\subsection{The action}

Let us now define the following generalized action 

\begin{equation}
S = S_{\mathrm PG} + \frac{1}{2}\sum_n\sum^3_{i=1}\bar\psi_n \eta_i (n)
\left( U_{n,i} \psi_{n+i} - U^\dagger_{n-i,i}\psi_{n-i}\right) + 
S_\tau(x, y), 
\label{gaction}
\end{equation}

\noindent
with

\begin{eqnarray}
S_{\tau}(x, y) &=& x \frac{1}{2} \sum_{n}\bar\psi_n \eta_0 (n)\left( U_{n,0}
\psi_{n+0} - U^\dagger_{n-0,0}\psi_{n-0}\right) \nonumber \\
&+& y \frac{1}{2} \sum_{n} \bar\psi_n \eta_0 (n)\left(  U_{n,0}
\psi_{n+0} + U^\dagger_{n-0,0}\psi_{n-0}\right)
\label{actiontxy}
\end{eqnarray}

\noindent
where $x$ and $y$ are two independent parameters. The QCD action is recovered 
from $S_\tau (x, y)$ by choosing $x= \cosh(\mu)$ and $y= \sinh(\mu)$.

Let us consider the generalized QCD action (\ref{gaction}). A natural question 
is: what should we expect for the phase diagram of this model in the $x, y$ 
plane?

We are going to conjecture a minimal phase diagram for this generalized model 
of QCD. To this end let us analyze first the $y= 0$ line. The point $x =1$ in 
this line would correspond to standard QCD at vanishing chemical potential. 
Now let be $L_t$ the temporal size of our lattice and assume we are in the 
scaling region but at a physical temperature $T$ lower than the deconfining 
critical temperature $T_c$. The point $x = 1$ in the $y = 0$ line will be in 
the confined phase. If we increase now the inverse gauge coupling $\beta$, 
the physical temperature increases and for $\beta$ values large enough the 
point $x= 1$ will be in the unconfined phase. This strongly suggests the 
presence of a phase transition point in the $y = 0$ line approaching the 
$x = 1$ point in this line by increasing $\beta$ and eventually crossing it 
for $\beta$ values large enough. 

\FIGURE{
\centerline{\includegraphics*[width=250pt,angle=270]{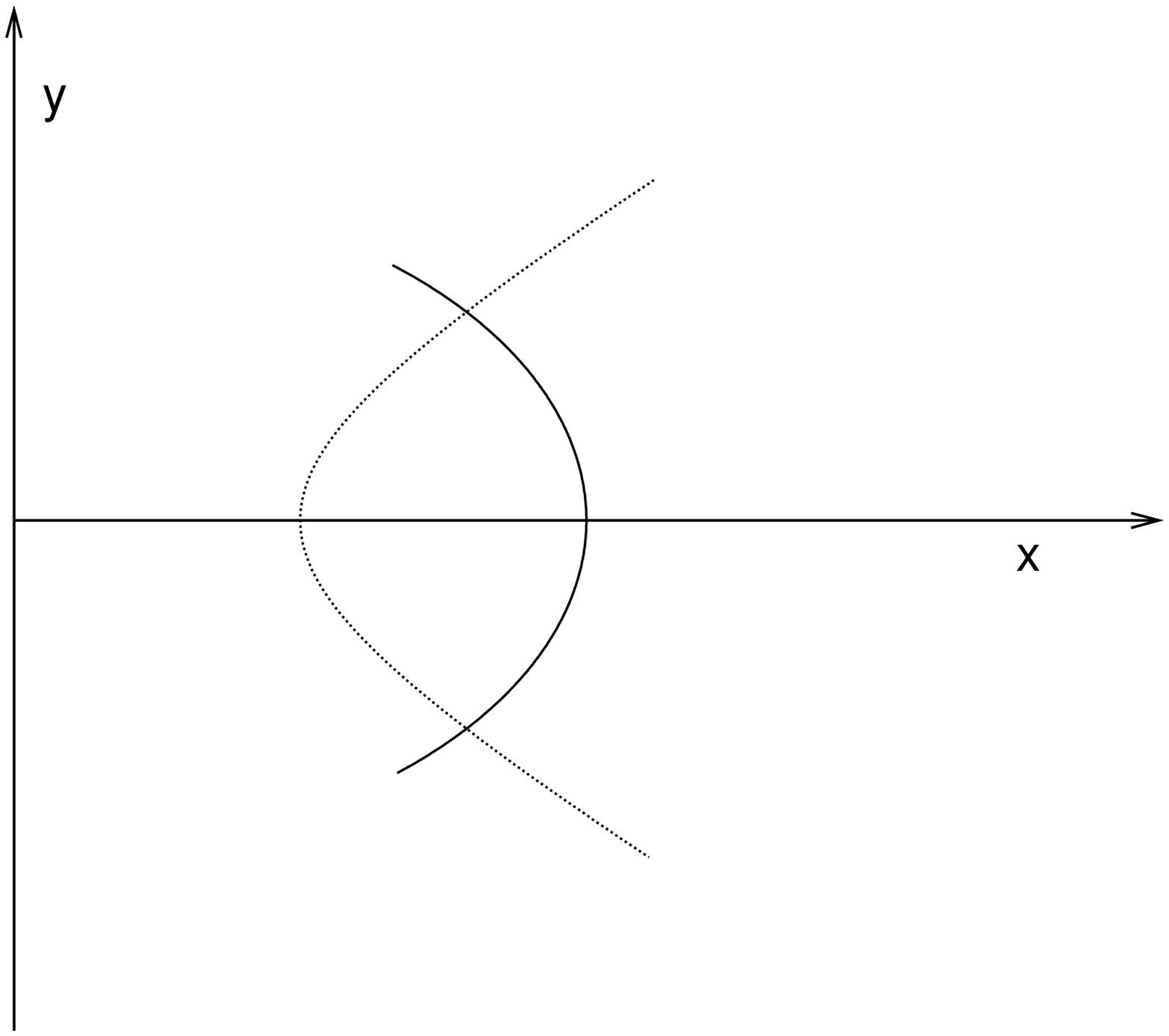}}
\caption{Minimal phase diagram conjectured for the generalized model of QCD.
The solid line is a line of phase transitions. The discontinuous 
line is the physical line $x^2 - y^2 = 1$.}
\label{fig1}
}

This argument can be enforced if we consider the strong coupling limit. 
In this limit, where the pure gauge action vanishes, one can change the 
temperature by using symmetric lattices and changing the temporal coupling 
in the fermion action. The point $x = 1$ in the $y = 0$ line would be zero 
lattice temperature whereas $x > 1$ would correspond to a temporal lattice 
spacing $a_\tau$ smaller than the lattice spacing $a$ in the space-like axis. 
Therefore $x > 1$ corresponds, in the strong coupling limit, to finite lattice 
temperature. But even in the strong coupling limit one reaches a chiral 
phase transition by increasing the lattice temperature \cite{strong}, 
thus suggesting 
that at $\beta = 0$ there is a phase transition point in the $y = 0$ line 
for a critical value of $x$, $x_c$, larger than 1. 

This argumentation drives one in a natural way to conjecture the minimal 
phase diagram shown in Fig. 1. The solid line would be a line of phase 
transitions that crosses the $y= 0$ axis at values of $x$ larger than 1 
for small $\beta$ values. By increasing $\beta$ and keeping fixed the 
temporal lattice extent $L_t$, the critical point on the 
$y= 0$ axis moves toward $x = 1$ and eventually crosses it. The discontinuous 
line in this figure stands for the physical line $x^2 - y^2 = 1$ along 
which one recovers standard QCD. Along this line 
points with a coordinate $y\ne 0$ would 
correspond to non vanishing chemical potential whereas $x=1, y=0$ is $\mu=0$. 
The intersection of the solid line with the discontinuous one will 
therefore give us 
the critical chemical potential of QCD at a given temperature. A 
change in the physical temperature can be simulated by changing $\beta$ 
keeping fixed $L_t$ or vice-versa. In both cases the solid line in Fig.1 
will move and the intersection point which gives the critical chemical 
potential will change with the physical temperature. The nature of the phase 
transition line will depend also on the other free parameters of the model 
i.e., number of flavors and quark masses.

\subsection{Symmetries of the partition function}

In order to get information on the symmetries of the partition function 
of the system described by the action (\ref{gaction}), let us 
define the following change of variables in the path integral for the 
Grassmann fields,

\begin{displaymath}
\psi^T_{jk} = e^{ik\alpha} \psi_{jk}\, , \hspace{1truecm}
\bar\psi^T_{jk} = e^{-ik\alpha} \bar\psi_{jk}\, ,
\end{displaymath}

\noindent
where $k$ is the temporal coordinate of a given site, $j$ stands for the three 
spatial coordinates, $e^{i\alpha}$ is an element of the group $Z_{L_t}$ and 
$L_t$ is the number of time slices. The only piece of the action 
(\ref{gaction}) not invariant under the previous variable change is 
$S_\tau (x, y)$, the different contributions of which transforming in the 
following way

\begin{eqnarray}
\left( \bar\psi_n \eta_0 (n) U_{n,0}\psi_{n+0}\right)^T 
&=& e^{\mathrm i\alpha} \left( \bar\psi_n \eta_0 (n) U_{n,0}\psi_{n+0}\right)
\nonumber \\
\left(\bar\psi_n \eta_0 (n) U^\dagger_{n-0,0}\psi_{n-0}\right)^T 
&=& e^{-\mathrm i\alpha} 
\left(\bar\psi_n \eta_0 (n) U^\dagger_{n-0,0}\psi_{n-0}\right)
\label{change}
\end{eqnarray}

In addition to the previous variable change for the Grassmann fields, we can 
also do a change of variables for the gluonic fields by multiplying each 
temporal link by an element $e^{i\gamma}$ of the center $Z_3$ of the $SU(3)$ 
group. Again the only non invariant contribution to (\ref{gaction}) is 
$S_\tau (x, y)$ and its different pieces transform in the same way as before, 
replacing $e^{i\alpha}$ in (\ref{change}) by $e^{i\gamma}$.

Using these transformations we can conclude that the integrated partition 
function of generalized QCD has the following symmetry 

\begin{equation}
Z_V\left(x+y, x-y\right) = 
Z_V\left( e^{\mathrm i\theta}(x+y), e^{-\mathrm i\theta}(x-y)\right), 
\label{symmetry}
\end{equation}

\noindent
where $e^{\mathrm i\theta}$ is now an element of $Z_{3L_t}$. 
Since the partition 
function must be independent of the parameters introduced by changing 
variables in the path integral, the only possibility is that it 
depends on $x, y$ through the combinations $(x+y)(x-y)$, $(x+y)^{3kL_t}$, 
$(x-y)^{3kL_t}$, with k any natural number. Taking also into account $CPT$ 
transformations we get that the dependence in the last two variables should 
appear as $(x+y)^{3kL_t}+(x-y)^{3kL_t}$, a result which could be also 
obtained by making the Polyakov loop expansion of the fermion determinant. 
Simple algebra tell us also that $(x+y)^{3kL_t}+(x-y)^{3kL_t}$ can be written 
as a combination of powers of $(x+y)^{3L_t}+(x-y)^{3L_t}$ and of $x^2-y^2$. 
Therefore we can say without loss of generality that the partition function 
$Z_V(x, y)$ will be, at finite volume,  an analytic function 
$\bar Z_V (u, v)$ of the two variables

\begin{eqnarray}
u &=& x^2 -y^2\, , \nonumber \\
v &=& (x+y)^{3L_t}+(x-y)^{3L_t}\, ,
\label{uv}
\end{eqnarray}

\noindent
and this is a non trivial information that we are incorporating in the 
description of the model. 

\FIGURE{
\centerline{\includegraphics*[width=250pt,height=400pt,angle=270]{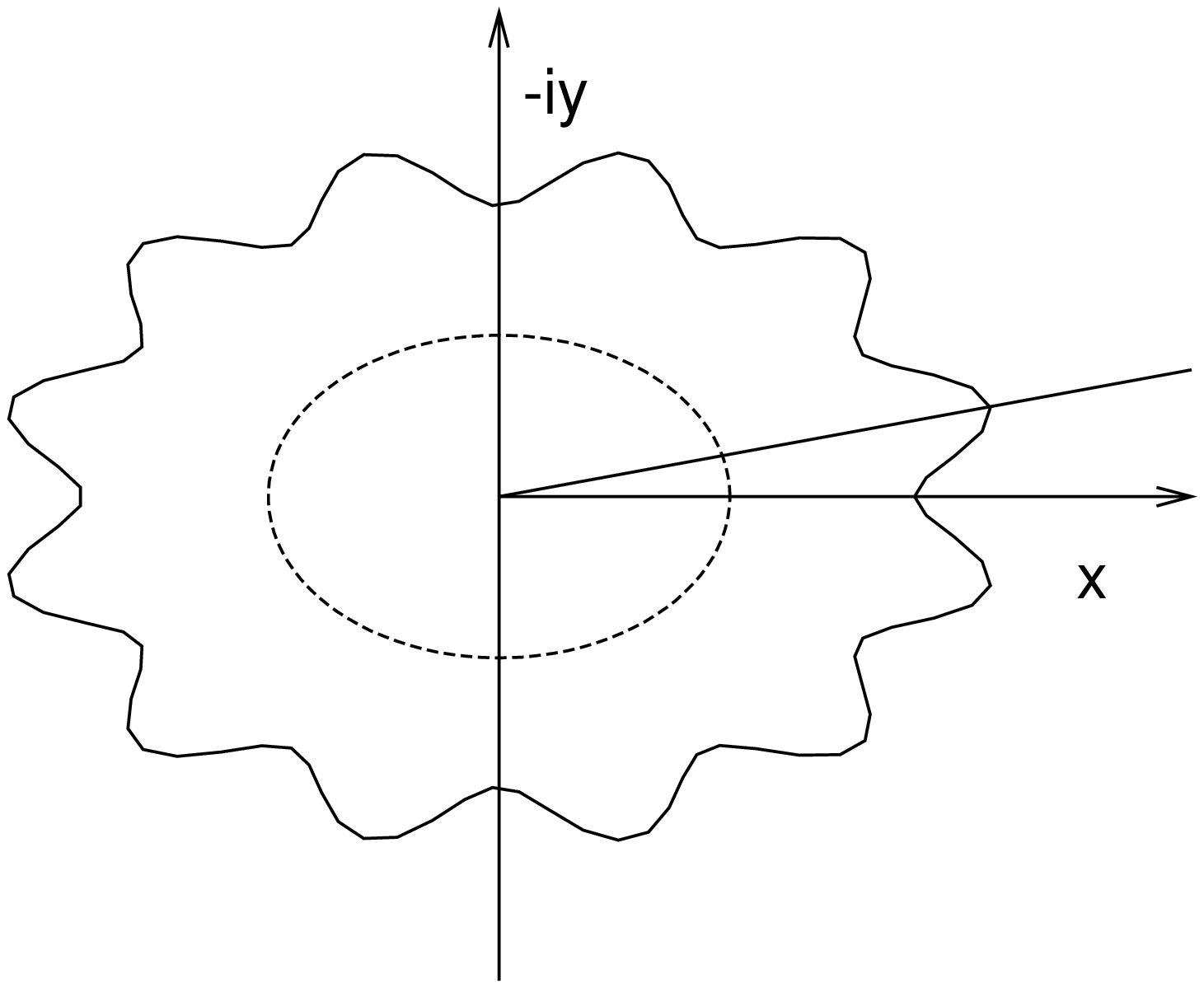}}
\caption{Conjectured phase diagram in the $x, \bar y$ plane. We have 
incorporated the property of periodicity.
The dashed line contains the only points accessible to numerical simulations 
of $QCD$ at imaginary chemical potential.}
\label{fig2}
}

\subsection{The extension to imaginary y}

The numerical analysis of the model (\ref{gaction}) for real values of 
$x, y$ is not possible since, as stated before, we find again the sign 
problem. However if we take the $y$ parameter as a pure imaginary number, 
$y= i\bar y$, $\bar y$ real, numerical simulations are feasible since the 
fermionic matrix, as for the case of imaginary chemical potential, is the 
sum of a constant diagonal matrix plus an antihermitian matrix, the last 
anticommuting with the $\gamma_5$ Dirac matrix. The eigenvalues of the 
fermion matrix appear then as pairs of complex conjugate numbers, and the 
fermion determinant is therefore real and positive.

Let be $z = x + i\bar y$ the complex number which we can associate to each 
pair of real values for $x, \bar y$. We can write $z$ as $z=\rho e^{i\eta}$ 
with 

\begin{equation}
\rho = \left(x^2 +\bar y^2\right)^{1/2}\, , 
\hspace{1truecm}
\tan\eta = \frac{\bar{y}}{x} \, .
\label{fase}
\end{equation}

The previous analysis on the symmetries of the partition function tell us 
that for imaginary values of the $y$ parameter, this function will depend 
only on $\rho$ and $\rho^{3L_t} \cos (3L_t \eta)$ i.e., the free energy will 
be a 
periodic function of $\eta$ with period equal to $2\pi/(3L_t)$. In 
particular if the phase transition line of Fig. 1 continues to imaginary 
values of $y$, the conjectured minimal phase diagram in the $x, \bar y$ plane 
would be that of Fig. 2, where we have incorporated the property of 
periodicity.

\FIGURE{
\centerline{\includegraphics*[width=180pt,angle=270]{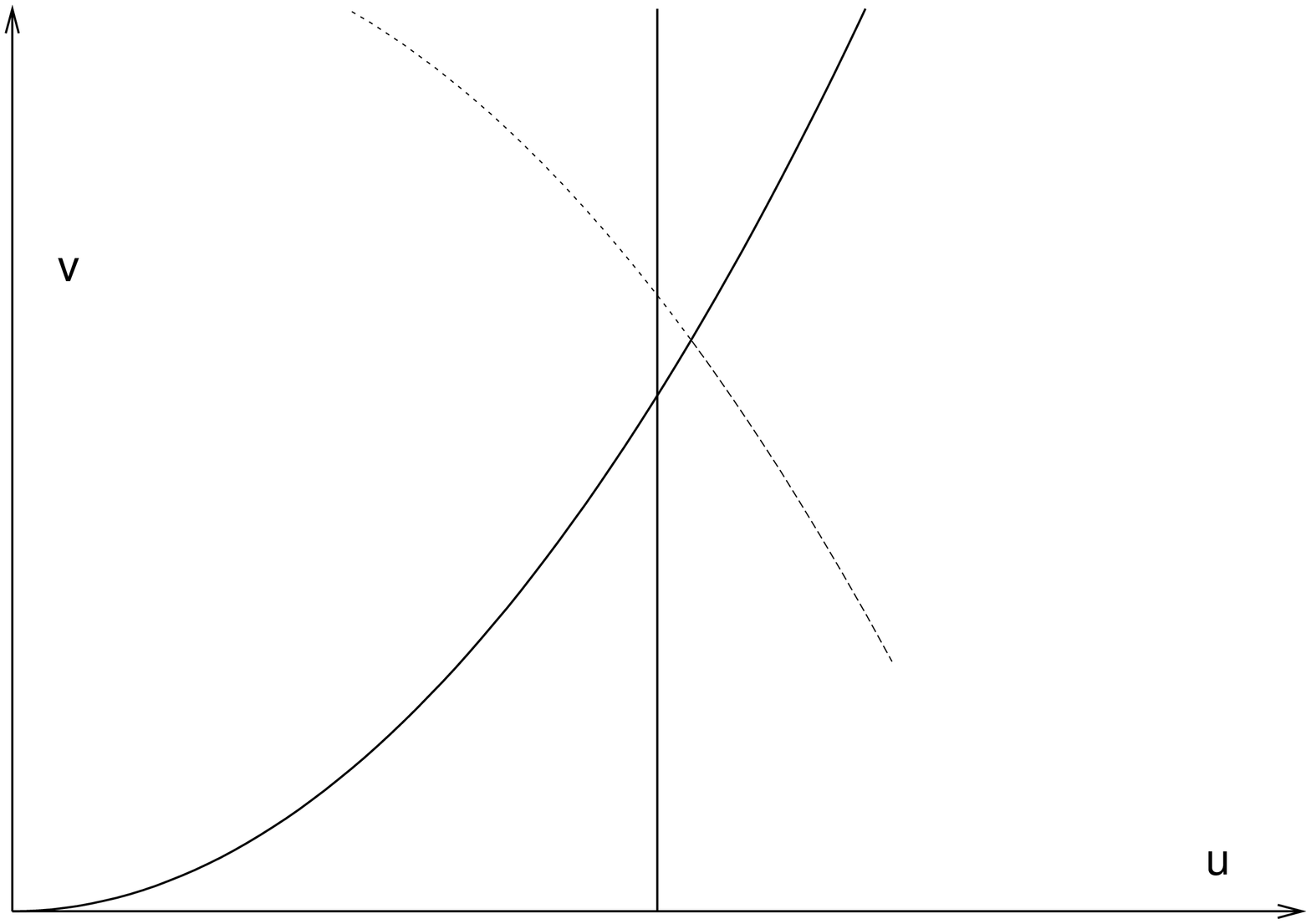}}
\caption{Phase diagram in the $(u,v)$ plane. The solid line 
$v = 2 u^{\frac{3}{2}L_t}$ corresponds to the $y = 0$ line of Fig. 1. 
The physical line $u = 1$ and the hypothetical phase 
transition line are also included.}
\label{fig3}
}

In Fig. 2 we have also included the line $\rho = 1$ (dotted line) which 
contains the only points accessible to numerical simulations of $QCD$ at 
imaginary chemical potential. As stated in the introduction, one can see 
now how this approach has in principle much more potentialities than the 
imaginary chemical potential approach. Indeed by increasing the inverse 
gauge coupling $\beta$, the phase transition line of Fig. 2 moves approaching 
more and more the origin of coordinates. In some interval $(\beta_m, \beta_M)$ 
the transition line intersects the $\rho=1$ line and then a phase transition 
will appear at imaginary chemical potential. In such a situation, the physical 
temperature is so high that the system is in an unconfined phase for any real 
value of the chemical potential. In order to understand something on the 
actual phase diagram from simulations at imaginary chemical 
potential, we need not 
only to extrapolate from imaginary to real values of $\mu$ but also to 
extrapolate from higher to lower temperatures or equivalently from higher to 
lower values of $\beta$. The advantage of our approach is that we can 
avoid the second kind of extrapolations because in our simulations 
$\rho$ is not enforced to be 1.

This discussion can become more clever if we go from the $x, \bar y$ plane to
the $u, v$ plane, where $u$ and $v$ are the variables defined in (\ref{uv}).
The reason for choosing this plane is that it contains simultaneously the two
planes of Figures 1 and 2. In Fig. 3 we have plotted the line

\begin{equation}
v = 2 u^{\frac{3}{2}L_t}
\label{y0}
\end{equation}

\noindent
which corresponds to the $y = 0$ line of Fig. 1. We have also included in
Fig. 3 the line $u = 1$ which contains the discontinuous or physical line
of Fig. 1 (points up the intersection point with the line 
$v = 2 u^{\frac{3}{2}L_t}$) as well as the dotted line of Fig. 2 
(points down the intersection point).
All the plane of Fig. 2 maps into the region down the line 
$v = 2 u^{\frac{3}{2}L_t}$ of Fig. 3, 
and this region is therefore accessible by simulating at
imaginary $y$.  On the other hand, the plane of Fig. 1 maps into the
region up the solid line of Fig. 3. The only points in
this Figure accessible to numerical
simulations at imaginary chemical potential are those on the line $u = 1$,
down the intersection point with the line $v = 2 u^{\frac{3}{2}L_t}$. 

In Fig. 3 we include also a hypothetical but plausible
realization of a phase 
transition line at a given physical temperature (simulations would be 
performed at fixed $\beta$ and $L_t$).

\subsection{Determination of the critical chemical potential near the 
continuum}

In order to get the critical value of the chemical potential we should 
determine the coordinates of the intersection point between the solid line 
and the physical line in Fig. 1. From the symmetries of the partition 
function we know that the phase transition line is an 
even function of $x$ and $y$. We 
can write therefore the following equation for the critical line of Fig. 1.

\begin{equation}
x^2 = 1 + a_0\left(\beta, L_t\right) + a_2\left(\beta, L_t\right) y^2 
+ a_4\left(\beta, L_t\right) y^4 
+ O(y^6), 
\label{criticalline}
\end{equation}

By fixing the lattice temporal extent $L_t$ and the gauge coupling $\beta$ 
one fixes the physical temperature $T$. The intersection point of the 
critical line with the physical line 

\begin{equation}
y^2_c = a_0\left(\beta, L_t\right) + a_2\left(\beta, L_t\right) y^2_c 
+ a_4\left(\beta, L_t\right) y^4_c 
+ O(y^6_c), 
\label{criticalmu}
\end{equation}

\noindent
will give us the critical value 
of the chemical potential ($y_c = \sinh(\mu_c a)$) at this 
temperature. 

If we approach the 
continuum limit by fixing the physical temperature $T$, i.e, by sending 
$\beta$ to $\infty$ keeping the product of $L_t$ times the lattice spacing 
$a$ fixed, we can write equation (\ref{criticalmu}) in the 
following way

\begin{eqnarray}
\left(\frac{\mu_c}{T_c}\right)^2 + \frac{1}{3} \frac{\mu^4_c}{T^2_c} a^2 
+ \cdots &=& 
a_0\left(\beta, L_t\right) L^2_t + a_2\left(\beta, L_t\right) 
\left(\frac{\mu_c}{T_c}\right)^2 
\nonumber \\
&+& \left[a_4\left(\beta, L_t\right) + \frac{1}{3} a_2
\left(\beta, L_t\right) \right]
L^{-2}_t \left(\frac{\mu_c}{T_c}\right)^4 + \ldots
\label{criticalmu2}
\end{eqnarray}

\noindent
where in order to get this equation we have multiplied both sides of 
(\ref{criticalmu}) by $L^2_t$. 
The coefficients $a_k\left(\beta, L_t\right)$ 
will depend in general on $L_t$ and on the dimensionless 
parameter $\Lambda/T$, where $\Lambda$ is the dimensional renormalization 
group parameter that fixes the QCD scale, and we have neglected for simplicity 
in all this analysis the dependence on the quark masses. Equation 
(\ref{criticalmu2}) then can be rewritten as follows

\begin{eqnarray} 
0 &=& \bar a_0\left(\frac{\Lambda}{T}, L_t\right) L^2_t + 
\left( \bar a_2\left(\frac{\Lambda}{T}, L_t\right) - 1 \right) 
\left(\frac{\mu_c}{T_c}\right)^2  
\nonumber \\
&+& \left[\bar a_4\left(\frac{\Lambda}{T}, L_t\right) + \frac{1}{3} \bar a_2
\left(\frac{\Lambda}{T}, L_t\right) - \frac{1}{3}  a^2 L^2_t \right]
L^{-2}_t \left(\frac{\mu_c}{T_c}\right)^4 + \ldots
\label{criticalmu3}
\end{eqnarray}

\noindent
which suggests the following scaling laws for the coefficients

\begin{equation}
\bar a_k\left(\frac{\Lambda}{T}, L_t\right) 
= c_k\left(\frac{\Lambda}{T}\right) \bar a_0
\left(\frac{\Lambda}{T}, L_t\right) L^k_t, 
\end{equation}

The critical value of $\mu/T$ will be given by the solution of the 
equation (\ref{criticalmu3}). This 
equation suggests that for $\mu_c/T_c$ small, the 
determination of the first two coefficients of (\ref{criticalline}) could 
be enough for the computation of the critical chemical potential. The goodness 
of this approximation will depend on the values of the coefficients of 
(\ref{criticalline}). In the next 
section we show that this approximation works quite well in the Gross-Neveu 
model, even at large values of $\mu_c/T_c$.

The strategy for the determination of the critical chemical potential is 
then the following. From numerical simulations at imaginary values of $y$, 
$y=i\bar y$, near the phase transition point $((1+a_0)^{1/2}, 0)$ 
one can locate several 
phase transition points in the $(x, \bar y)$ plane (see Fig. 2). By fitting 
these points with equation (\ref{criticalline}) with the + sign of the 
coefficient proportional to $y^2$ replaced by -, we can numerically measure 
the first coefficients. The critical value of the chemical potential, 
$\mu_c$, will then be given by

\begin{equation}
\mu_c = \pm \sinh^{-1} \left(\frac{a_0}{1-a_2}\right)^{1/2}.
\label{muc}
\end{equation}

An alternative approach for the determination of the critical chemical 
potential is to use the variables $(u, v)$ instead of $(x, y)$. Since near 
the continuum $u$ takes values near 1, the phase transition line of Fig. 3 
could be fitted with the following equation

\begin{equation}
v = b_0 + b_1 \left(u-1\right) +  b_2 \left(u-1\right)^2 + 
O\left(\left(u-1\right)^3\right).
\label{criticalline2}
\end{equation}

\noindent
The coefficient $b_0$ fully determines the critical $\mu$ through the 
following equation

\begin{equation}
\mu_c = \frac{1}{3L_t} \cosh^{-1} \left(\frac{b_0}{2}\right).
\label{muc2}
\end{equation}

The use of the $(u,v)$ variables incorporates information on 
the symmetries of the free energy density. Furthermore consistency between 
results obtained from fits with equations (\ref{criticalline}) and 
(\ref{criticalline2}) would improve the confidence level on the 
full procedure.

\subsection{Two other alternative ways for extracting the critical $\mu$: 
the $(\beta,y)$ and $(\beta,x)$ planes}

\FIGURE{
\centerline{\includegraphics*[width=200pt,angle=270]{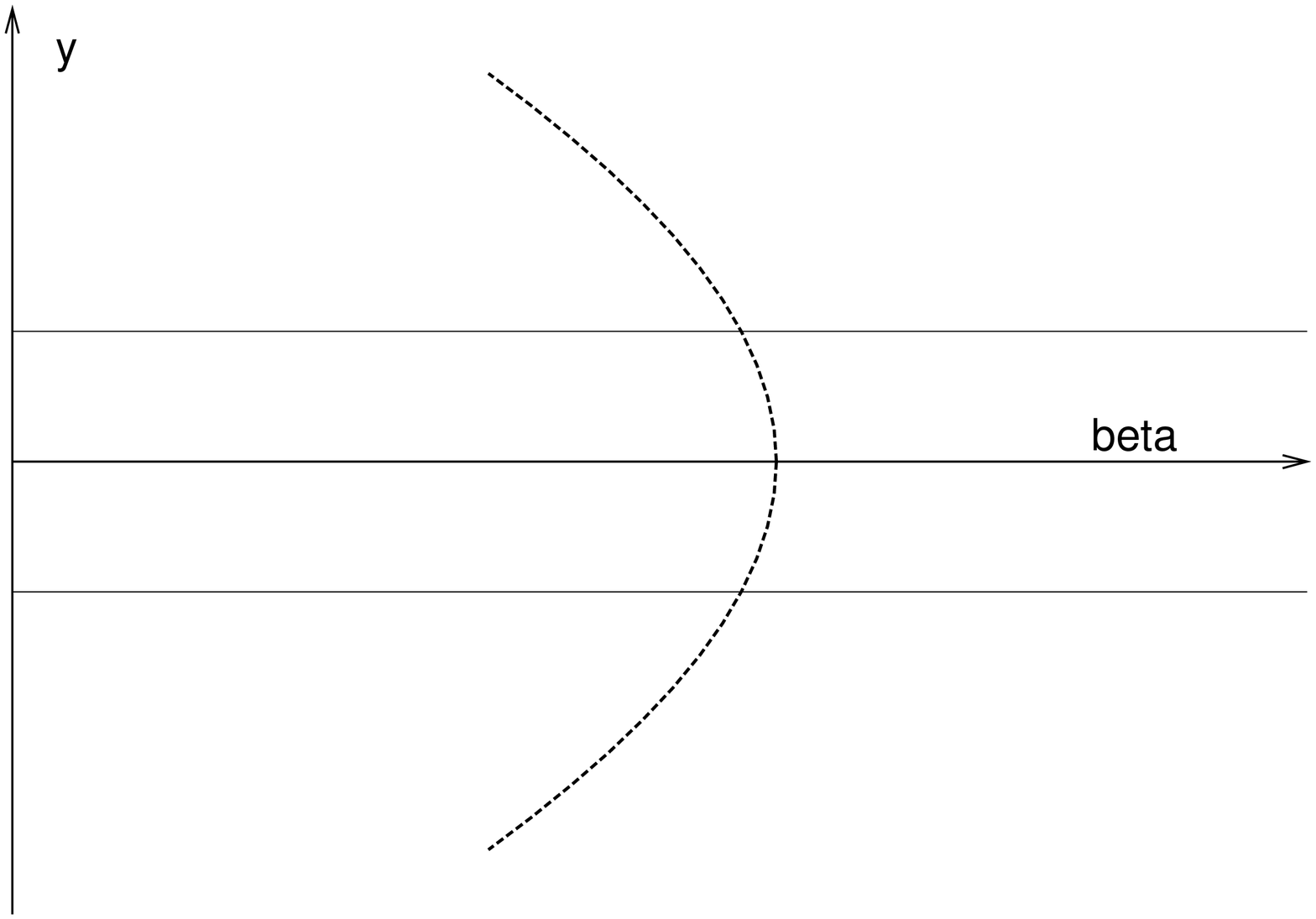}}
\caption{Conjectured phase diagram in the $(\beta,y)$ plane. 
The solid line is the physical line whereas the discontinuous one 
is a line of phase transitions. }
\label{fig4}
}

In the previous paragraph it was assumed that the inverse gauge coupling 
$\beta$, the quark masses and the lattice temporal size $L_t$ were fixed 
and then we explored the phase diagram in the $(x,y)$ plane. In physical 
terms this means that, if $\beta$ is large enough to be in the scaling 
window, we fix the physical temperature $T$ and the physical quark masses. 
However it could be interesting in practice to analyze the projection of 
the phase diagram into the other two planes $(\beta,y)$ and $(\beta,x)$ for 
the following two reasons:
\begin{itemize}
\item[\it i.] it would give an independent determination of $(\mu_c, T_c)$;
\item[\it ii.] one could apply some powerful techniques in the numerical 
analysis, as the Ferrenberg-Swendsen approach \cite{ferrenberg}.
\end{itemize}
Let us start with the case in which we fix $x=x_0>1$, $L_t$ and the lattice 
quark masses. The conjectured phase diagram in the $(\beta,y)$ plane is 
reported in Fig. 4. The solid line is the physical line

\begin{displaymath}
y = y_0 = \sinh\left(\cosh^{-1} (x_0)\right),
\end{displaymath}

\noindent
and the discontinuous line is a line of phase transitions. 
Using again the symmetries 
of the partition function we can write for this line the following equation 

\begin{equation}
\beta = \beta_0\left(x_0, L_t\right)  + b_2 \left(x_0, L_t\right) y^2 
+ O\left(y^4\right),
\label{clineybeta0}
\end{equation}

\noindent
where we have neglected for simplicity the dependence of the coefficients on 
the quark masses.

We are interested in the determination of the intersection point between the 
physical and phase transition lines of Fig. 4. 
As in the case discussed in the previous section, one 
can measure $\beta_0$ and $b_2$ from 
simulations at $y=0$ and at imaginary $y=i\bar y$ (keeping $x_0$ fixed). 
Once we estimate the lattice spacing $a$, assumed we are in 
the scaling region, 
we can obtain $\beta_c$, $\mu_c$ and the critical values of the physical 
quark masses. Notice also that in the imaginary chemical potential approach 
$x_0$ is enforced to be less than 1.

The last case to analyze, the projection of the phase diagram into the 
$(\beta, x)$ plane, is less interesting from a practical point of view 
because:

\noindent
{\it i}. The physically relevant region in the $(\beta,x)$ plane (keeping 
$y=y_0$ fixed) is $x$ larger but near to 1. The equation for the phase 
transition line near $x=1$ can therefore be written as

\begin{equation}
\beta = \beta_0\left(y_0\right) + a_1\left(y_0\right) \left(x-1\right) 
+ a_2\left(y_0\right) \left(x-1\right)^2 
+ O\left(\left(x-1\right)^3\right).
\label{criticalbetax}
\end{equation}

\noindent
The critical inverse gauge coupling $\beta_c$ will be obtained by putting 
$x=x_0$ in (\ref{criticalbetax}), with 

\begin{displaymath}
x_0 = \cosh \left(\sinh^{-1} (y_0)\right).
\end{displaymath}

\noindent
But since no symmetry enforces now $a_1(y_0)$ to vanish, we should do, in 
contrast with the previous cases, a three parameters fit to get $\beta_c$ 
to order $(x_0-1)^2$.

\noindent
{\it ii}. Numerical simulations must be performed at 
imaginary values of $y$ and 
therefore we need, at the end of the procedure, to do also an extrapolation 
of $\beta_c(y)$ from imaginary values of $y$ to real $y=y_0$.

\subsection{Vacuum expectation values on the phase transition line}

In order to get a deep understanding of the phase transition(s) at finite 
baryon density we would like to know not only the value of the critical 
chemical potential at a given physical temperature but also how vacuum 
expectation values of physical quantities change when crossing the phase 
transition line. To this end we will focus here on two typical observables, 
the mean plaquette energy and chiral condensate, and on the first theory 
scheme described previously i.e., the analysis of the phase diagram in the 
$x, y$ plane at fixed $\beta$. The generalization to other observables as 
the baryon number density, and, or to other projections of the 
three-dimensional phase diagram will be straightforward.

The chiral condensate $q(x^2, y^2)$ and plaquette energy $E(x^2, y^2)$ at 
fixed $\beta$ will be analytical even functions of $x$ and $y$ in all the 
points very near to the phase transition line both on the left and on the 
right of it. If we compute for instance the chiral condensate $q(x^2, y^2)$ 
in all the points of the phase transition line 

\begin{displaymath}
x^2 = 1 + a_0\left(\beta, L_t\right) + a_2\left(\beta, L_t\right) y^2 
+ a_4\left(\beta, L_t\right) y^4 
+ O(y^6)\, , 
\end{displaymath}

\noindent
approaching it from the left-hand side, we expect an analytical even function 
of $y$. Then we can write for the chiral condensate along this line the 
equation

\begin{equation}
q_{c-}\left(y^2 \right) = q^{(0)}_{c-} + q^{(2)}_{c-} y^2 
+ O_-\left(y^4\right), 
\label{condensate1}
\end{equation}

\noindent
and the same for all the points on the right of the phase transition line 

\begin{equation}
q_{c+}\left(y^2 \right) = q^{(0)}_{c+} + q^{(2)}_{c+} y^2 
+ O_+\left(y^4\right). 
\label{condensate2}
\end{equation}

From simulations at imaginary $y = i\bar y$ one can measure the chiral 
condensate in both sides along the phase transition line, and by fitting the 
numerical results with these equations, we would get a numerical determination 
of the parameters in (\ref{condensate1}), (\ref{condensate2}). This procedure 
would allow us to extend the chiral condensate results to the physical point 
$y= y_c$.

\section{The Gross-Neveu model at finite density}

To check all the conjectures and ideas developed in section 2, we decided 
to use as a toy model the four-Fermi model in three dimensions with discrete 
chiral symmetry at finite $\mu$ and $T$. The continuum Lagrangian is

\begin{equation}
L = \sum^{N_f}_{j=1} \left[ \bar\psi^{(j)}\dsl\psi^{(j)} 
- \frac{g^2}{2N_f}\left(\bar\psi^{(j)}\psi^{(j)}\right)^2\right]\, ,
\label{grossneveu}
\end{equation}

\noindent
where $\psi^{(j)}$ is a four-component spinor, $N_f$ is the number of flavors
and the discrete $Z_2$ symmetry is $\psi\rightarrow \gamma_5 \psi$, 
$\bar\psi\rightarrow -\bar\psi\gamma_5$. This model has a non trivial 
renormalization group fixed point characterized by the spontaneous breaking 
of the chiral symmetry at a finite critical coupling $g^2_c$, its $1/N_f$ 
expansion about the fixed point $g^2_c$ being renormalizable. In addition 
the model can be solved analytically in the large $N_f$ limit and then we 
can compare numerical predictions with exact results in this limit. Finally 
the phase structure of the model in the $(\mu, T)$ plane resembles to 
that expected for QCD, with a phase transition line separating a 
phase where chiral symmetry is spontaneously broken from a phase where 
the symmetry is restored.

After the introduction of an auxiliary scalar field $\sigma$, equation 
(\ref{grossneveu}) becomes

\begin{equation}
L = \sum^{N_f}_{j=1} \left[ \bar\psi^{(j)}\dsl\psi^{(j)} + 
\sigma \bar\psi^{(j)}\psi^{(j)}\right]
+ \frac{N_f}{2g^2}\sigma^2\, , 
\label{grossneveu2}
\end{equation}

\noindent
the vacuum expectation value of $\sigma$ being a good order parameter for the 
chiral symmetry restoration transition.

Regularized in a space-time lattice with staggered fermions, the Gross-Neveu 
action is \cite{sacha}

\begin{eqnarray}
S &=& \sum^{N_f/2}_{i=1}\sum_n\sum^3_{\mu=1} \frac{1}{2}\eta_{\mu}(n) 
\bar\psi_i(n)\left[\psi_i(n+\mu) - \psi_i(n-\mu)\right]
\nonumber \\ 
&+& \frac{1}{8} 
\sum^{N_f/2}_{i=1} \sum_n \bar\psi_i(n)\psi_i(n) \sum_{<\bar n, n>} 
\sigma(\bar n)
\:+\: \frac{N_f}{4g^2}\sum_{\bar n}\sigma^2(\bar n)\, , 
\label{grossneveul}
\end{eqnarray}

\noindent
where the symbol $<\bar n, n>$ denotes the set of eight dual lattice  sites 
$\bar n$ surrounding the lattice site $n$, $\eta_0(n)=1$ and 
$\eta_{\mu}(n) = (-1)^{n_{\mu}+n_{\mu-1}}$ for $\mu=1,2$.

The inclusion of a chemical potential $\mu$ selecting a finite fermion density 
is achieved, as in QCD, by multiplying all forward (backward) temporal links 
by $e^{\mu}$, $(e^{-\mu})$. The fermionic bilinear part of the action then 
becomes

\begin{eqnarray}
S_F &=& \sum^{N_f/2}_{i=1}\sum_n \frac{1}{2} \bar\psi_i(n) \left[e^{\mu} 
\psi_i(n+0) - e^{-\mu} \psi_i(n-0)\right] 
\nonumber \\
&+& \sum^{N_f/2}_{i=1}\sum_n\sum^2_{\alpha=1} \frac{1}{2} 
\bar\psi_i(n)\eta_{\alpha}(n)
\left[\psi_i(n+\alpha) - \psi_i(n-\alpha)\right] 
\nonumber \\
&+& \frac{1}{8} \sum^{N_f/2}_{i=1}  
\sum_n \bar\psi_i(n)\psi_i(n) \sum_{<\bar n, n>} \sigma(\bar n)
\label{grossneveuf}
\end{eqnarray}

\FIGURE[!h]{
\centerline{\includegraphics*[width=180pt,angle=270]{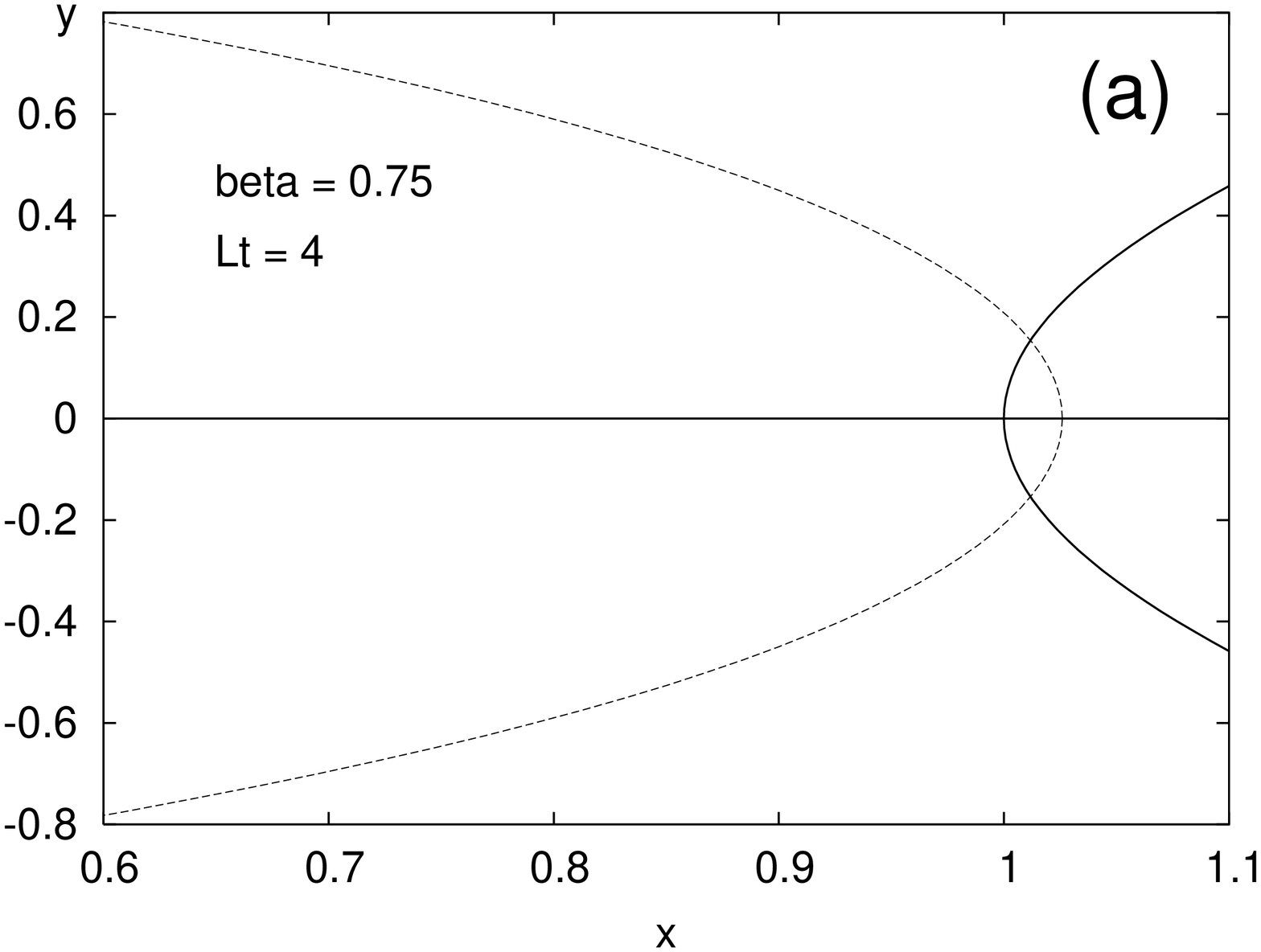}}
\centerline{\includegraphics*[width=180pt,angle=270]{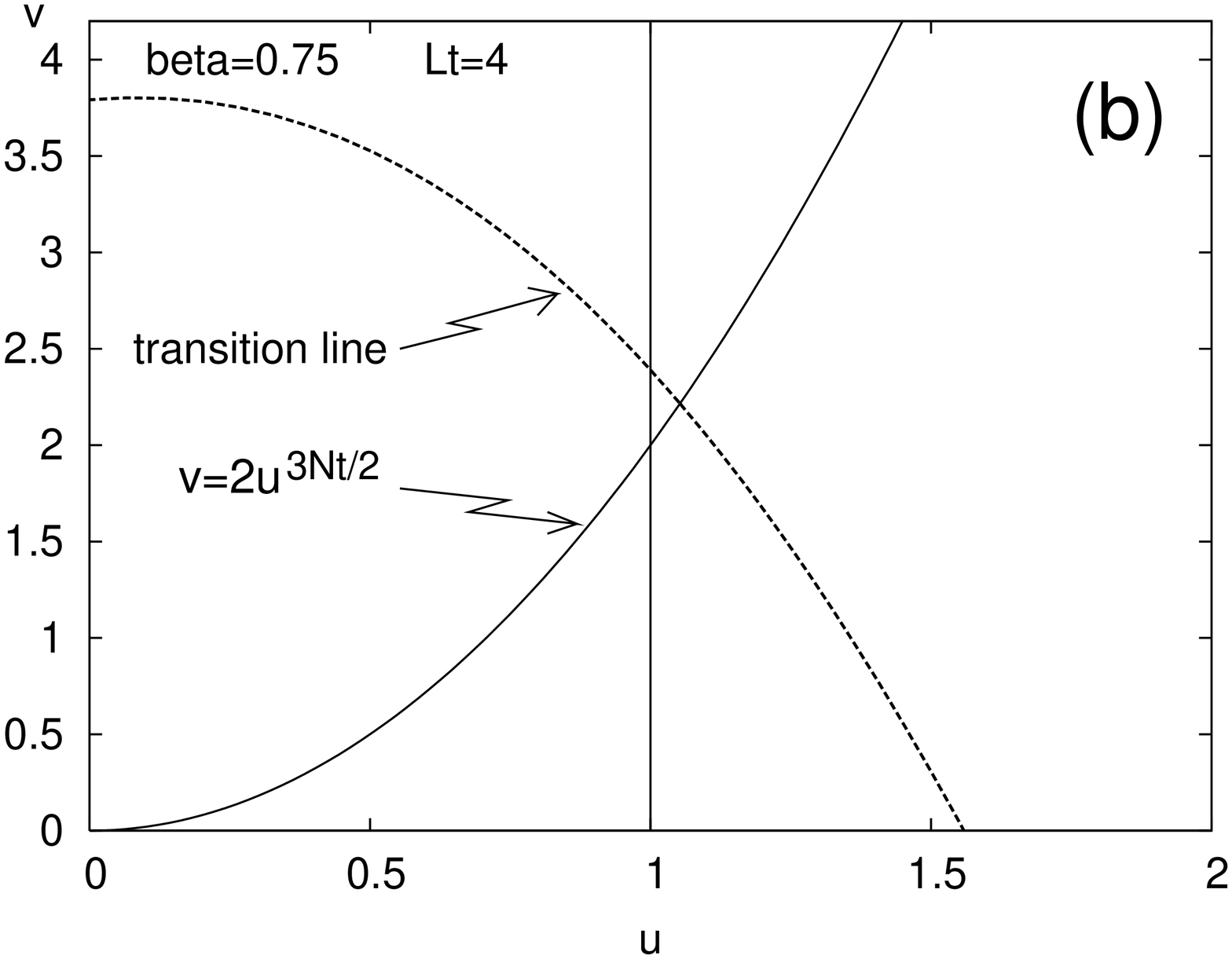}}
\centerline{\includegraphics*[width=180pt,angle=270]{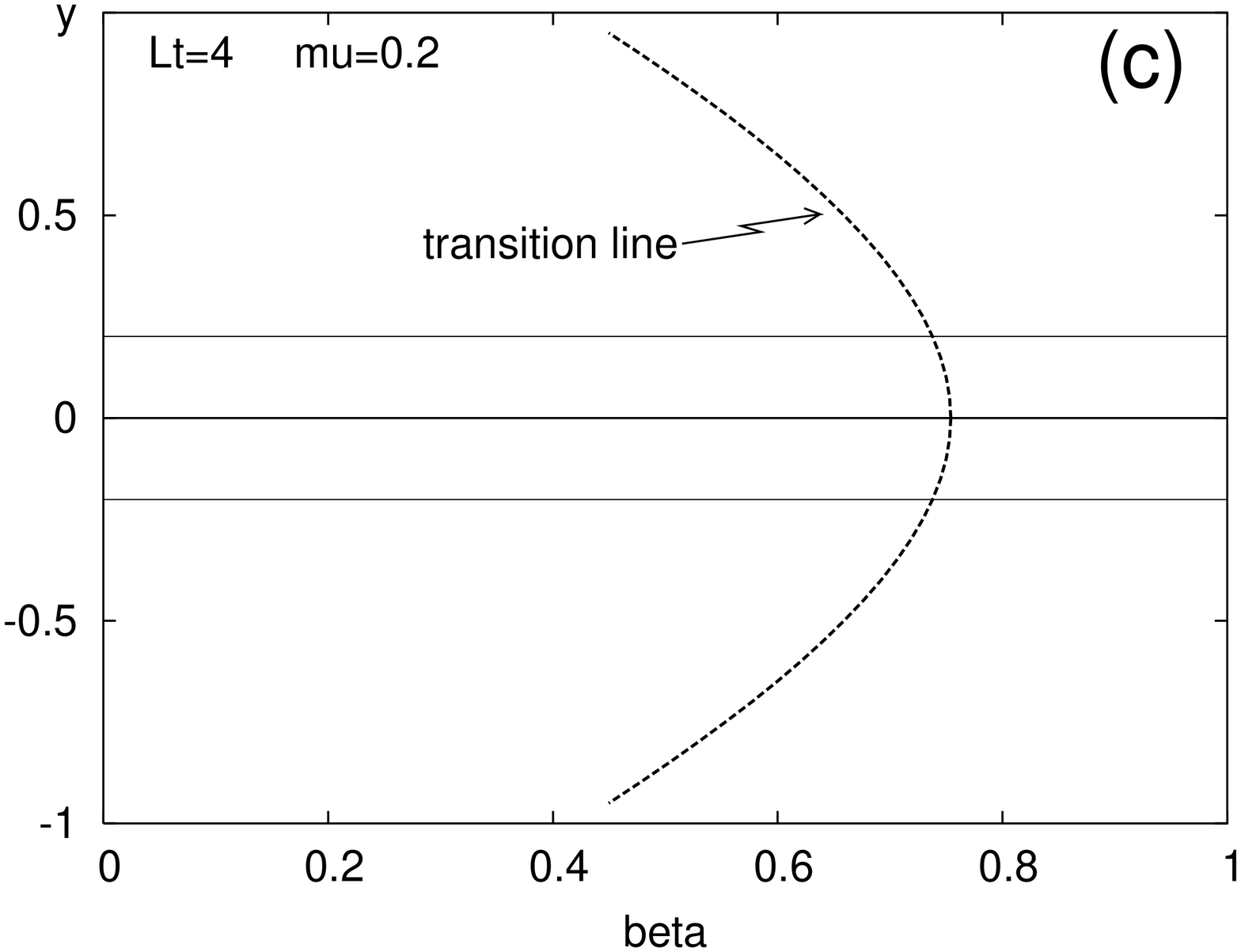}}
\caption{ Phase diagrams of the generalized four-Fermi model in the 
$(x, y)$ (5a), 
$(u, v)$ (5b) and $(1/g^2,y)$ (5c) planes respectively. 
All the phase diagrams 
correspond to a lattice temperature $1/L_t=1/4$ in lattice units. The 
inverse four-Fermi coupling $1/g^2=\beta =0.75$ in Figs. 5a and 5b whereas 
$x=1.0211$ in Fig. 5c. 
The solid line is the physical line whereas the 
discontinuous line is a second order phase transition line.}
\label{fig5abc}
}

\clearpage

In the large number of flavors limit, $N_f\rightarrow\infty$, fluctuations 
around the saddle point solution are suppressed and then the vacuum 
expectation value $\Sigma$ of the scalar field $\sigma$ can be 
self-consistently determined using the gap equation 

\begin{equation}
\Sigma = -g^2 \left<\bar\psi\psi\right>, 
\label{gap}
\end{equation}

\noindent
where $\left<\bar\psi\psi\right>$ here means the mean value of the chiral 
condensate in the free fermion theory at finite chemical potential and with 
an effective fermion mass $m=\Sigma$.

The formalism developed in the previous section can be straightforwardly 
applied here. Indeed we can define a generalized fermion action

\begin{eqnarray}
S_F(x, y) &=& 
x \frac{1}{2}\sum^{N_f/2}_{i=1} \sum_n \bar\psi_i(n)\left[
\psi_i(n+0) - \psi_i(n-0)\right] 
\nonumber \\
&+& y \frac{1}{2} \sum^{N_f/2}_{i=1} \sum_n \bar\psi_i(n)
\left[\psi_i(n+0) + \psi_i(n-0)\right] 
\nonumber \\
&+& \sum^{N_f/2}_{i=1}\sum_n\sum^2_{\alpha=1} \frac{1}{2} 
\bar\psi_i(n) \eta_{\alpha}(n)
\left[\psi_i(n+\alpha) - \psi_i(n-\alpha)\right] 
\nonumber \\
&+& \frac{1}{8} \sum^{N_f/2}_{i=1}
\sum_n \bar\psi_i(n)\psi_i(n) \sum_{<\bar n, n>} \sigma(\bar n)\, ,
\label{grossneveufg}
\end{eqnarray}

\noindent
that agrees with action (\ref{grossneveuf}) for $x=\cosh(\mu)$, 
$y=\sinh(\mu)$. In the 
large $N_f$ limit the generalized model can also be analytically solved, and 
the solution is again given by the gap equation (\ref{gap}) where now 
$\left<\bar\psi\psi\right>$ means the mean value of the chiral condensate 
in the generalized free-fermion model at an effective fermion mass $m=\Sigma$. 

We are now in good conditions to check if the conjectures formulated in 
section 2 are realized in the four-fermion model. 
Figures 5a, 5b and 5c show 
the phase diagrams of the generalized four-Fermi model in the $(x, y)$, 
$(u, v)$ and $(y, g^2)$ planes respectively. All the phase diagrams 
correspond to a lattice temperature $1/L_t=1/4$ in lattice units. The 
inverse four-Fermi coupling $\beta=1/g^2=0.75$ in Figs. 5a and 5b whereas 
$x=1.0211$ ($\mu=0.2$) in Fig. 5c. 
A second order phase transition line separating a broken from an unbroken 
phase and crossing the physical line $x=\cosh\mu$, $y=\sinh\mu$ was found in 
all the cases, thus confirming actually all our conjectures.

To have a guess on the potentialities of our approach when applied to QCD we 
have done a further check. This check consisted in fitting several phase 
transition points in the $(x, y)$ plane at imaginary $y$ with 
equation (\ref{criticalline}) in the quadratic approximation (two-parameter 
fits). The 
purpose of the check was to assume these phase transition points 
as they were 
obtained from numerical simulations at imaginary $y$, where the sign problem 
is absent, and then to check the degree of precision with which we can 
predict $\mu_c$ and $T_c$.

\FIGURE{
\centerline{\includegraphics*[width=200pt,angle=270]{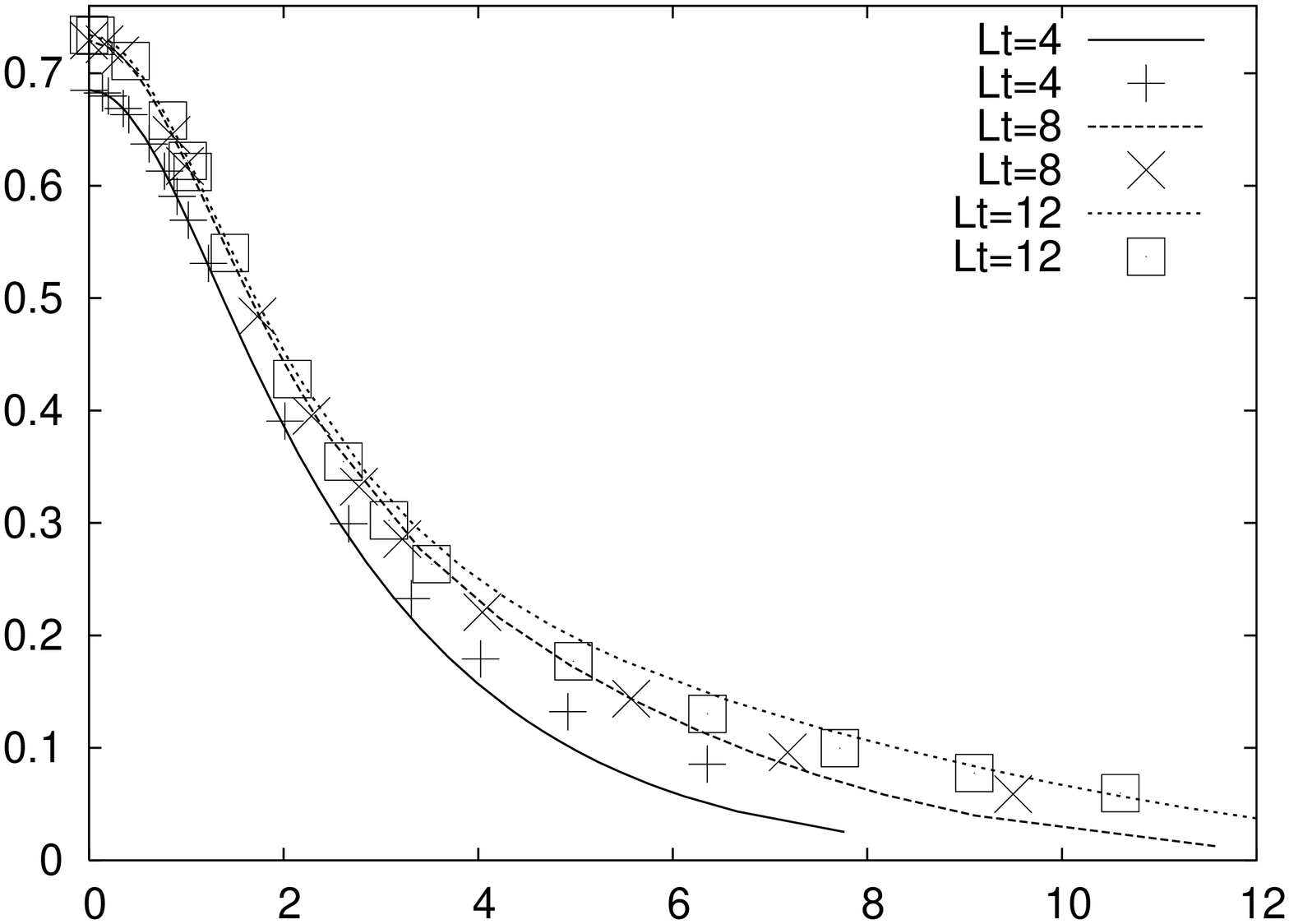}}
\caption{ Phase 
diagram in the $(\mu/T, T/\Sigma_0)$ plane for several values of the lattice 
temporal extent $L_t$. $\Sigma_0$ is the expectation value of the scalar field 
$\sigma$ in the ground state at $T=0$. 
The lines correspond to analytical results at the same
values of $L_t$ while the symbols correspond to the 
extrapolated results.}
\label{fig6}
}

To locate the physical region in which the agreement between the extrapolated 
results and the exact ones is better, we have plotted in Fig. 6 the phase 
diagram in the $(\mu/T, T/\Sigma_0)$ plane for several values of the lattice 
temporal extent $L_t$. $\Sigma_0$ is the expectation value of the scalar field 
$\sigma$ in the ground state at $T=0$, and the plots have been done assuming 
asymptotic scaling. The lines correspond to analytical results. 
The observed dependence on $L_t$, especially at large 
$\mu/T$, shows that the effects of regularization are still important in 
this region for these lattices. One can also observe in this figure how the 
extrapolated results agree very well with the exact ones even at large 
values of $\mu/T$. 

Finally and in order to confront our method with the imaginary chemical 
potential approach, we have plotted in Fig. 7 the phase diagram of the 
Gross-Neveu model in the $\mu ,1/g^2$ plane as obtained from the analysis 
of the $x, y$ plane at imaginary $y$ in the quadratic approximation of 
equation (\ref{criticalline}) 
(crosses), and from the imaginary chemical potential approach (dashed lines), 
the last obtained from quadratic as well as quartic fits of the critical 
coupling as a function of the imaginary chemical potential. 
The solid line stands for
the exact results on a lattice of temporal extent $L_t=4$. As can be seen
both approaches agree with the analytical results in the small chemical
potential region, but results obtained with our method follow quite well
the exact ones in the large $\mu$ region in contrast with the imaginary
chemical potential results that give even negative values for $g^2_c$
at large $\mu$ in the quadratic approximation. 
From both theoretical and practical points of view there is little hope 
to get reliable values for more than two-three coefficients from 
imaginary $\mu$ numerical data, so we improved the fit procedure, as 
stated before, considering up to quartic fits of the critical coupling as 
a function of the imaginary chemical potential.
As can be seen in Fig. 7 the extra degree of freedom improves the
results at small $\mu$, but not in a relevant way.

\FIGURE{
\centerline{\includegraphics*[width=200pt,angle=270]{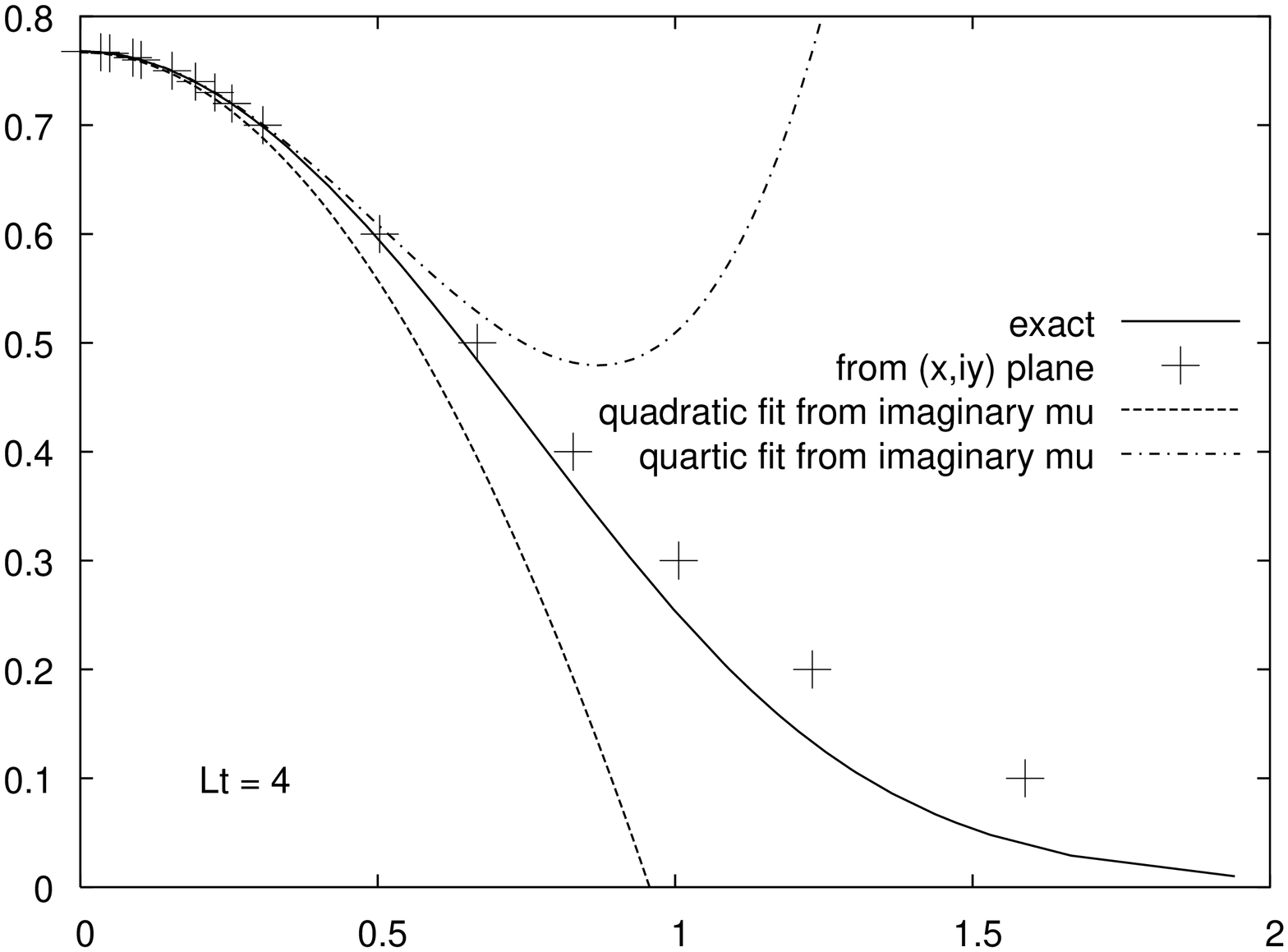}}
\caption{ Phase diagram of the
Gross-Neveu model in the $ \mu, 1/ g^2$ plane as obtained from the analysis
of the $x, y$ plane at imaginary $y$ (crosses), and from the imaginary 
chemical potential approach (dashed line). The solid line stands for
the exact result on a lattice of temporal extent $L_t=4$}
\label{fig7}
}

\section{Four flavor QCD at finite T and $\mu$}

Understanding the finite temperature and finite density phase structure of 
QCD is one of the main challenges of lattice field theory. Whereas qualitative 
and quantitative progress on the phase structure of the model at vanishing 
baryon density has been achieved in the last years, the sign problem has 
prevented, as discussed in the introduction of this paper, any significant 
progress on the understanding of the behavior of matter at high 
density. 

We want 
to devote this section to the application of the approach described in the 
previous sections to the analysis of the phase diagram of four flavor QCD 
at finite baryon density. The aim of such an analysis is twofold: first we 
want to verify if the conjectures formulated in the first section, and 
corroborated in the previous section for the Gross-Neveu model, do actually 
work in QCD, and second we want to gain some experience in order to 
analyze in a future work the more interesting cases of 2 or 2+1 flavor 
QCD.

Even if a good understanding of the phase diagram of the four flavor model 
would be of interest, the aim of this section is not to do an exhaustive 
analysis of the four flavor model but rather to verify if the approach 
reported in this paper can hopefully be applied to QCD. This is the reason 
why we have concentrated our 
efforts on the analysis of one point of the phase diagram, but using two 
of the independent approaches described in section 2: the determination of 
the critical values of $\mu$ and $T$ from the analysis of the phase diagrams 
in the $(x, y)$ and $(\beta, y)$ planes. Since the two approaches are based on 
independent numerical simulations and independent fits, the agreement 
between the predicted critical parameters should be considered a proof
of the reliability of the results.

\subsection{The ($x, y$) plane}

The model analyzed is therefore that described by action (\ref{action}) i.e., 
QCD at finite $\mu$ and four degenerate staggered flavors. In order to 
determine a point of the phase transition line in the $(T, \mu)$ plane we 
performed HMC simulations of the generalized model described by action 
(\ref{gaction}), (\ref{actiontxy}) at several values of $x$ and imaginary 
values of $y= i\bar y$. The lattice size was $8^3\times 4$, the inverse gauge 
coupling $\beta= 5.05$, and the 
quark masses were fixed to $ma = 0.06$ in lattice units.
The corresponding physical temperature is $T \approx 136$MeV.

We have performed simulations for $7$ values of $\bar y$ from $0$ to $0.15$;
for each value of $y$ $8-13$ values of $x$ have been simulated with
statistics ranging from $10000$ to $50000$ HMC trajectories (the larger
ensembles near the critical points). The simulations have been performed on the
Linux clusters of LNGS-INFN, using a total of 50 CPUs.

\FIGURE{
\centerline{\includegraphics*[width=250pt,angle=270]{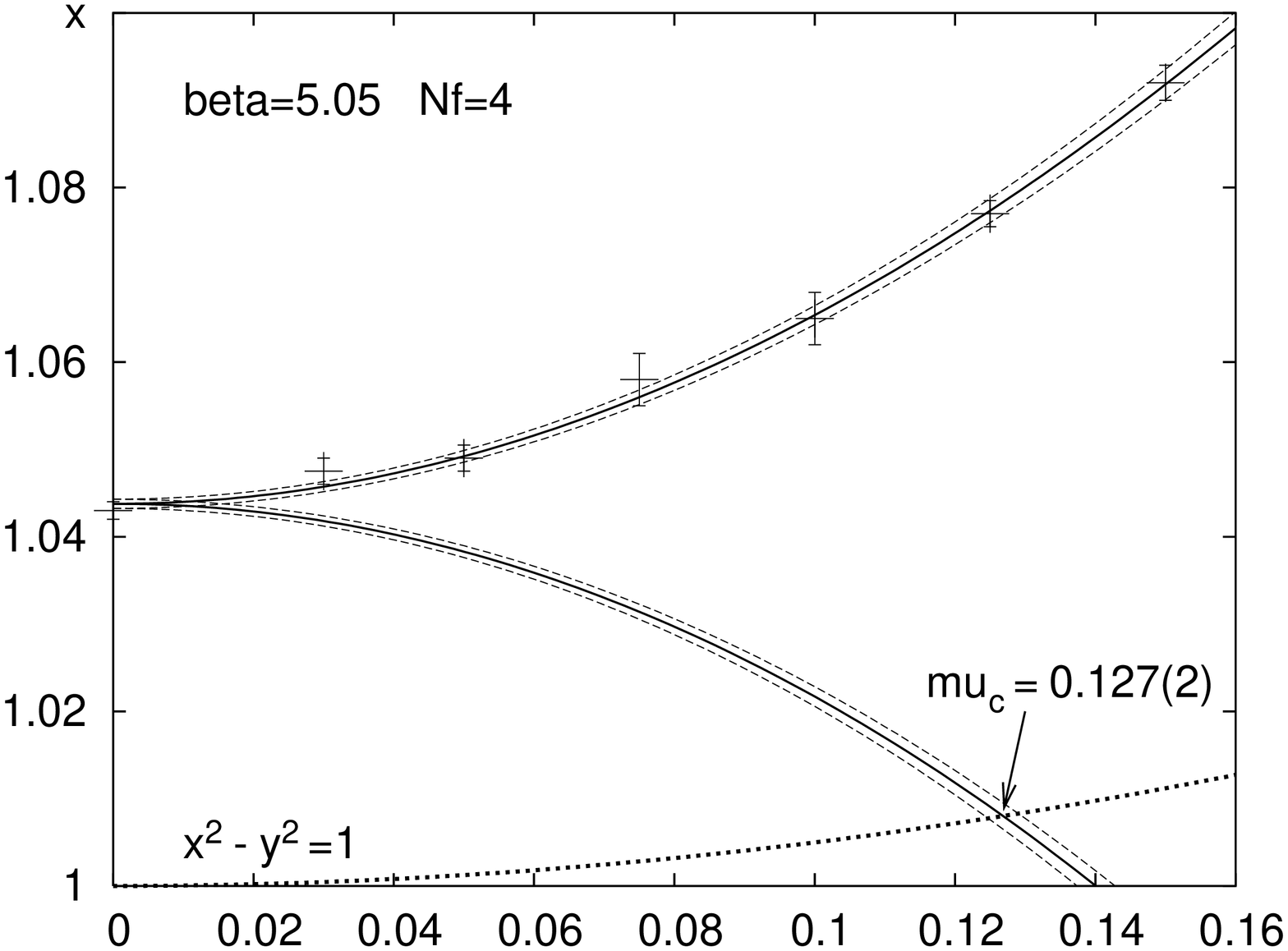}}
\caption{Measured critical points in the $(\bar y,x)$ plane at $\beta=5.05$ 
and $ma=0.06$. The upper and lower half figure correspond to imaginary and 
real $y$ respectively. The solid lines are a quadratic fit of the points and 
its analytical continuation to real $y$. We include also the 
physical line $x^2-y^2=1$. Dashed lines represent maximum deviation
of the fitting function within statistical errors.}
\label{fig8}
}

From these numerical simulations we have determined seven critical points 
in the $(\bar y,x)$ plane (see Fig. 8),
looking for the position of the maximum of the plaquette, chiral, 
and Polyakov loop susceptibilities. The position of 
the maximum was compatible for the three observables in all the cases, but 
the results for the plaquette susceptibility seem to be of better quality. 
The two-state 
signal observed for the three measured quantities in all the points plotted 
in Fig. 8 indicates a first order phase transition line. Unfortunately we 
have no data on larger lattices which would allow a finite size scaling 
analysis to confirm these expectations.

By fitting the seven points of Fig. 8 with a quadratic function (equation 
(\ref{criticalline}) with $a_n=0$ for $n>2$) we get the solid line in this 
Figure. Its analytical continuation to real values of $y$ has also been 
plotted in Fig. 8 where one can also observe the intersection point with the 
physical line $x^2-y^2=1$. The coordinates of this intersection point 
$x_c=1.0081(2), y_c=0.127(2)$ allow us to predict that the critical chemical 
potential in lattice units is $\mu_c=0.127(2)$ at $\beta=5.05$ and $ma=0.06$. 

We can fix the lattice scale from the mass of the $\rho$ resonance, 770 MeV.
In such case the results of reference \cite{scale} imply that the 
lattice spacing
is approximately $0.3$ fm at $\beta=5.2$ for $ma=0.06$, and we 
set the lattice spacing
at smaller values of $\beta$ by using the two loop $\beta$ function.
The phase transition at $\mu=0$ is located at $\beta\approx 5.08$
\cite{ahasen}. This gives $T_{\mathrm C}\approx 136$ MeV. Our results
reported in the previous paragraph give another point of the phase
transition line: $T_{\mathrm C}\approx 133$ MeV, 
$\mu_{\mathrm C}\approx 68(1)$ MeV.


\FIGURE{
\centerline{\includegraphics*[width=250pt,angle=270]{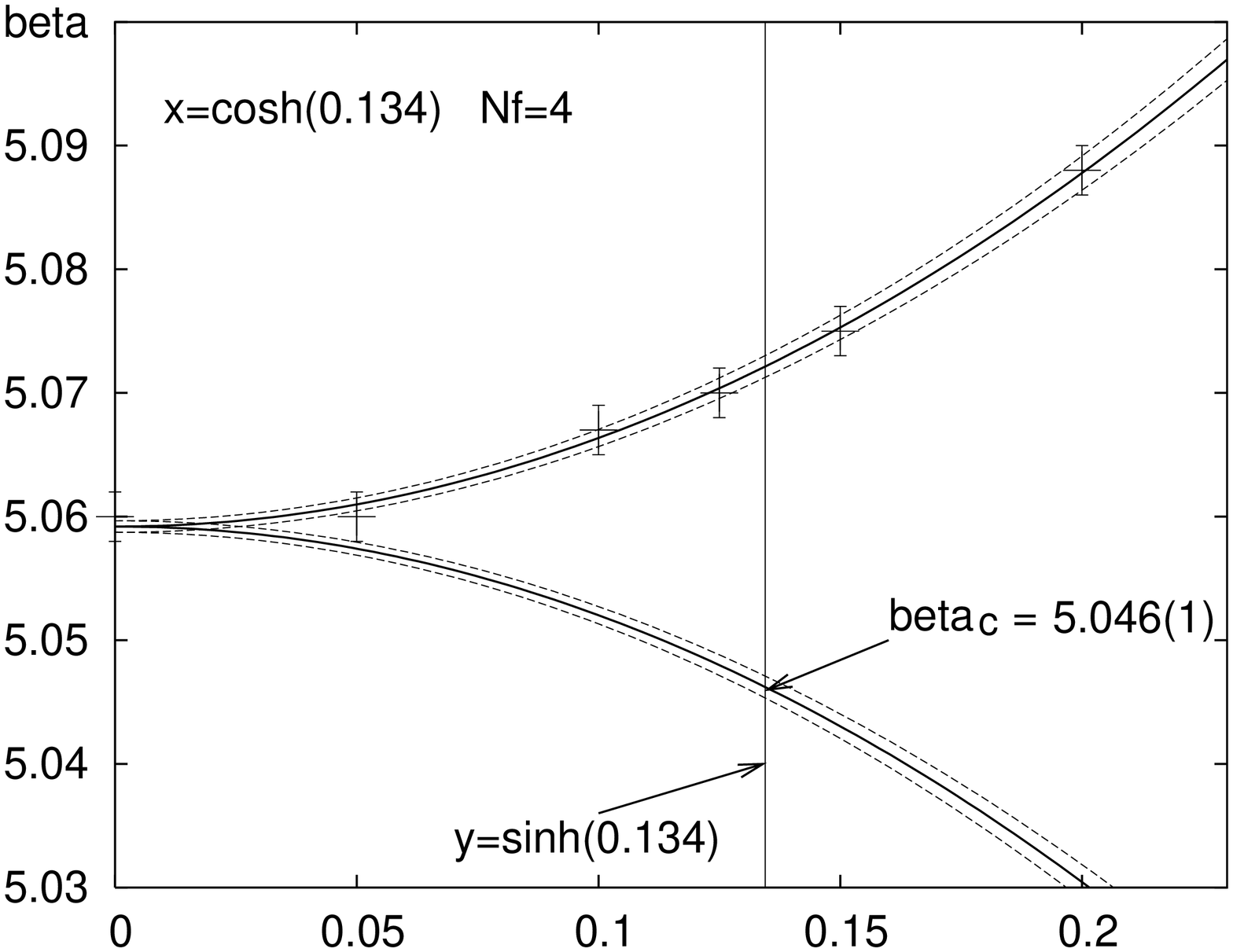}}
\caption{Measured critical points in the $(\bar y,\beta)$ plane at $x=1.009$ 
and $ma=0.06$. The upper and lower half figure correspond to imaginary and 
real $y$ respectively. The solid lines are a quadratic fit of the points and 
its analytical continuation to real $y$. We include also the 
physical line $y=\sinh(\cosh^{-1}(1.009))$. 
Dashed lines represent maximum deviation
of the fitting function within statistical errors.}
\label{fig9}
}

Once we have determined the coordinates of the critical point, we can apply 
the formalism described in section 2.6 to analyze the nature of the physical 
phase transition. A very interesting 
feature that emerges from our simulations is that the latent heat is
independent of the parameters $x,y$ within errors in such a way 
that there is not a real
need to perform an analytical continuation; therefore the physical phase 
transition is also first order. Similar results have also been obtained for 
the chiral condensate and Polyakov loop.

\subsection{The ($\beta, y$) plane}

By fixing $x=1.009$ (corresponding to $\mu=0.134$, our first 
rough estimation of 
the critical value at $\beta=5.05$ and $ma=0.06$), 
we have also performed simulations in the $(\beta, \bar y)$ plane,
using statistics and number of values for the parameters
$\beta$ and $\bar y$ similar to those of the $(x,y)$ case.
The aim of this calculation, as stated before, 
is to check if we get results consistent with those of section 4.1. 

Fig. 9 is the same as Fig. 8 but in the $(y,\beta)$ instead of the $(y,x)$ 
plane. The solid lines are a fit of the measured phase transition points 
with equation (\ref{clineybeta0}) and its analytical 
continuation to real values of $y$. Again in this case a clear two state 
signal was observed for the three measured observables. As in Fig. 8, we have 
also plotted in Fig. 9 the physical line that in this case is the line 
$y =\sinh(\cosh^{-1}(1.009))$. The intersection point of the 
phase transition line 
with the physical line gives the critical value $\beta_c=5.046(1)$ in close 
agreement with the results reported in the previous section, if
we take in mind that the value of $\mu$ used (0.134) is slightly larger 
than the actual critical $\mu$ at $\beta=5.05$ (0.127).
For this case the critical point is $T_{\mathrm C}\approx 132$ MeV, 
$\mu_{\mathrm C}\approx 71$ MeV.
Also for this set of simulations the global behavior of latent heat suggests
a first order transition.

\section{Conclusions and comments}

We have introduced a new approach to analyze the phase diagram of QCD at 
finite chemical potential and temperature, and applied it to 
the four-fermion model in three dimensions and to QCD with four
degenerate quarks.

Our scheme is based on the introduction of a generalized QCD action which 
depends on two extra free parameters, $x, y$. When the point $(x, y)$ 
belongs to the curve $x^2 - y^2 = 1$ one recovers standard QCD at finite 
chemical potential.

The generalized QCD action suffers also, for real values of $y$, from the 
sign problem. However at imaginary values of $y$ the fermion determinant is 
positive definite and therefore standard numerical simulations are feasible.

Even if our theory scheme can resemble the imaginary chemical potential 
approach, we have shown that its range of applicability is much wider. Indeed 
our approach shares with the imaginary chemical potential method the
need of analytical. However whereas imaginary 
chemical potential simulations have to be performed at high temperature
(large $\beta$) i.e., in the deconfined phase for real values of $\mu$, in 
our scheme simulations can be performed at any temperature and any value of 
$\mu/T$. The price to pay for that is one more parameter in the numerical 
procedure. 

To make evident the potentialities of our approach we have confronted it 
with the imaginary chemical potential method when applied to the Gross-Neveu 
model in the limit of large number of flavors, where analytical results are 
available. We have shown how both approaches agree with the analytical 
results in the small chemical potential region, but results obtained with 
our method follow quite well the exact ones in the large $\mu$ region in 
contrast with the imaginary chemical potential results, that give even 
negative values for $g^2_c$ at large $\mu$. An open point however is to 
dispose of a good criterion to understand in which region of the parameters 
space the quadratic approximation of equation (\ref{criticalline}) (or 
equivalently of equation (\ref{criticalmu3})) works well. For what concerns 
the Gross-Neveu model, the aforementioned region is actually very wide.

We have also performed simulations of QCD with four degenerate flavors
in order to test the procedure in a real QCD simulation. We have determined
the critical chemical potential for a value of $\beta$ inside the
scaling region and, using the second scheme depicted in section 2.5,
we found the critical $\beta$ for a given $\mu$ value, obtaining a
consistent pair of results.

As stated before, our theory scheme requires more computer time than other 
more standard simulations since we have one more parameter in the action. 
Roughly speaking the computer time needed to get a single point of the 
phase diagram in the $(\mu, T)$ plane is of the order of the one needed 
to get the critical line within the imaginary chemical potential approach. 
However, our approach, as we have shown for the Gross-Neveu model, seems 
able to reproduce quite well the phase diagram at values of $\mu/T$ not 
accesible to the other standard methods, and this is in our opinion the 
reason that makes it relevant for the field despite of its computing cost.

Concerning the Glasgow reweighting procedure in two
parameters space \cite{fodor}, one could consider the possibility of using 
the fermion determinant at the critical point of the $y=0$ line instead of the
determinant at $\mu=0$ in the integration reweighted measure. 
We suspect that hereby one could improve the overlaps.

Finally we plan for near future to apply our approach to the analysis of 
two-color QCD at finite $\mu$ and to 2 and 2+1 flavor QCD.

\acknowledgments

This work has been partially supported by an INFN-CICyT collaboration
and by Ministerio de Ciencia y Tecnolog\'{\i}a (Spain), projects
FPA2003--02948 and BFM2003--08532--C03--01/FISI.
V.L. has been supported by Ministerio de Ciencia y Tecnolog\'{\i}a 
(Spain) under the Ram\'on y Cajal program.

\end{document}